\documentclass[aps,pre,preprint]{revtex4-1}

\usepackage{epsf}
\usepackage{latexsym}
\usepackage{epsfig}
\usepackage{amsmath}
\usepackage{amsfonts}
\usepackage{amssymb}
\usepackage{graphicx}
\usepackage{color}
\usepackage[usenames,dvipsnames]{xcolor}

\newcommand{\prs}[1]{{\left(#1\right)}}
\newcommand{\col}[1]{{\left[#1\right]}}
\newcommand{\chs}[1]{{\left\{#1\right\}}}
\newcommand{\prob}[1]{{\mathcal{P}\prs{#1}}}
\newcommand{\vc}[1]{{\boldsymbol #1}}
\newcommand{\avg}[2]{{\left<#1\right>_{#2}}}
\newcommand{\bu}{{\bar{u}}}
\newcommand{\bw}{{\bar{w}}}
\newcommand{\erf}{{\mbox{erf}}}
\newcommand{\sgn}{{\mbox{sgn}}}
\newcommand{\bC}{{\bar{C}}}
\newcommand{\cN}{{\mathcal{N}}}

\begin{document}

\title{Non-Thermal Transitions in $n$-th Order Moral Decisions}
\author{Roberto C. Alamino}
\affiliation{Non-linearity and Complexity Research Group, Aston University, Birmingham B4 7ET, UK}

\begin{abstract}
This work introduces a model in which agents of a network act upon one another according to three different kinds of moral decisions. These decisions are based on an increasing level of sophistication in the empathy capacity of the agent, a hierarchy which we name \emph{Piaget's Ladder}. The decision strategy of the agents is non-rational, in the sense that it does not minimize model's Hamiltonian, and the model presents quenched disorder given by the distribution of its defining parameters. We obtain an analytical solution for this model in the thermodynamic limit and also a leading order correction for finite sized systems. Using these results, we show that typical realizations develop a rich phase structure with discontinuous non-thermal transitions.
\end{abstract}

\maketitle

\section{Introduction}

Human societies are inherently complex. So much that recent research indicates that even highly specialized qualitative knowledge from social sciences does not improve prediction performance on average \cite{Tetlock15}. On the other hand, statistical physics models have been successful in reproducing and predicting observed emergent behavior in real social datasets \cite{Castellano09, Abrams11, Vicente14, Ross15}, giving birth to a whole new branch of physics, still in its infancy, which has been named \emph{Sociophysics} \cite{Galam12}.   

Social scientists and psychologists often see these results with suspicion, arguing that each human individual is different. It is however clear that human behavior presents identifiable patterns without which their very disciplines would not exist. The difficulty comes from the numerous sources of disorder which can affect individual behavior and their decisions concerning social interactions \cite{Reichardt11, Martinez14} as, for instance, the influence of mass media \cite{Clementi15} or the availability of natural resources \cite{Sugiarto15}. The contribution of statistical physics has been in understanding how emergent behaviors depend on the disorder and which are the relevant parameters driving transitions between the different phases \cite{Sole11}.  

One of the most studied problems in sociology concerns the emergence of moral behavior and its consequences to the general well-being and survival ability of a certain population \cite{Wilson03}. There is evidence that moral behavior is the result of multi-level selection \cite{Okasha06, Schonmann12}, but the exact mechanism is still not understood. Although the picture is not yet complete, some of its parts are gradually becoming clearer. For instance, the essential role of intuitive (emotional) judgments in the process of moral formation and decision making is now well accepted and has been recently used in the formulation of the framework known as Moral Foundations Theory (MFT) \cite{Haidt07}.  
 
The essential role of emotions is, in fact, nothing but expected as motivation from emotional satisfaction, independently of cultural differences, is a factor that has been identified a long time ago as being decisive in the survival of an organism or a group even when their physical needs are fulfilled \cite{Maslow43}. 

In this work we use a simplified statistical mechanics model to study the influence of different moral decisions in the overall emotional satisfaction of a human population. We will not be interested in the problem of emergence of particular moral beliefs and, therefore, we will assume that the concepts of what is morally acceptable is agreed \emph{a priori} by the whole population. It is obvious that these concepts will vary and even exchange places in different societies, but this will not affect our analysis.

The model analyzed here is infinite dimensional in the sense that each person from the group interacts with every other person by taking a one-time binary decision: to act \emph{helpfully} or \emph{harmfully} relative to the group's agreed moral rules. The person's decision is assumed to be a \emph{conscious} one, which is why we can actually call it a moral decision. We then classify the moral decisions according to the level of empathy embodied in them.  

Although MFT suggests that humans classify moral beliefs using at least five dimensions instead of only one binary dimension (harmful/helpful), this simplification is enough for the purposes of this work. A more sophisticated model can be obtained by considering the agents as neural networks classifying moral multidimensional binary vectors as in \cite{Vicente14}. 

The spectrum of moral decisions that can be taken by humans can be extremely complicated and analyzing it would be out of the scope of this paper. Here we focus on what we call $n$-th order moral decisions, where the order of the  decision identifies the increasing level of empathy necessary to take it according to the work of the well-known psychologist Jean Piaget \cite{Piaget65}. Piaget observed that there are distinct cognitive stages in the development of the child intellect before it reaches the adult stage. This evolution occurs in three steps with increasing levels of empathic capacity. It is this sequence of three steps that we call \emph{Piaget's Ladder} and which defines the order of a moral decision. 

Piaget has proposed, based on empirical observations, that every children first develops a sense of self in which it is capable of understand its own feelings, but is unable to recognize the feelings of others, acting only selfishly. Accordingly, we call this step the Selfish Step and decisions taken selfishly are considered of 0th order. As its development proceeds, the child becomes capable of recognizing that others also have feelings, but it is still unable to see things from others' perspective. We call it the Parental Step (1st order) as it is not uncommon to parents to project their desires in the way the act towards their children. Finally, the cognitive abilities of the child reach a stage in which it can finally understand that others have different needs. This is the Empathic Step (2nd order). The details of Piaget's original theory of cognitive development have been several times revised to account for further experimental evidence \cite{Oakley04}, but the details  will not be as important for our purposes, only its key idea of incremental empathy.

The agents of our network will act on one another using moral decisions which are combinations of these three steps. These decisions, although conscious, are not rational in the sense that agents do not devise strategies to maximize their well-being (to be defined rigorously later) and simply follow a limited set of pre-determined rules. It is well known that in a varied number of scenarios, this is usually the rule rather than the exception \cite{Ariely08}. The average well-being of the group defines then the phases of the statistical physics model and we are then interested in characterizing how they change as the set of disorder parameters of the model is varied.  

A detailed explanation of how the model is constructed will be given in Sec. \ref{section:TSM} together with its analytical solution. The phase structure of this solution is obtained and analyzed in Sec. \ref{section:PS}. Finally, we present our conclusions and further discussion in Sec. \ref{section:Conc}.

\section{The Model}
\label{section:TSM}

\begin{figure}
  \centering
  \includegraphics[width=8cm]{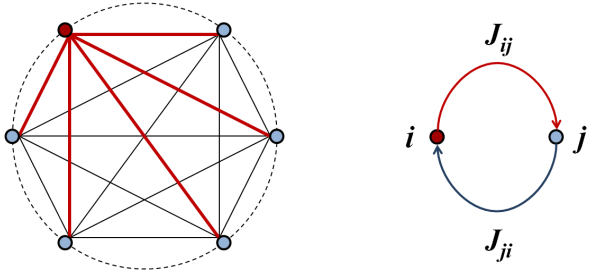}
  \caption{Fully connected social network (left) with non-symmetrical interactions between agents (right).}
  \label{figure:FCN}
\end{figure}

We consider a population of $N$ agents represented by the nodes of a fully connected social network, as the one shown in fig.\ref{figure:FCN} (left). Links between nodes indicate direct interactions between agents. As the aim of this work is primarily to identify the emergence of collective behavior in general, we solve only the fully connected case. The model can be straightforwardly generalized to any topology, with an obvious increased difficulty for obtaining exact solutions and we therefore will leave their analysis to forthcoming papers.

In the thermodynamic limit of $N\to\infty$ (in which the fully connected case flows to the mean-field solution) the model develops a very rich phase structure with different kinds of non-thermal transitions. These transitions are induced by varying the intensity of the disorder, understood to be the parameters of the different probability distributions of the model which we will describe in the following.

Each static configuration of the network is defined by one single (frozen) realization of three variables, namely $J$, $\vc{u}$ and $\vc{v}$. The variable $J$ is a matrix containing the moral decisions taken by the agents towards others. A certain realization of $J$ is obtained by each agent on a generic site $i$ acting on another agent at site $j$ by taking a one-time binary decision: it either acts \emph{harmfully} or \emph{helpfully} towards $j$, where these notions are judged according to \emph{a priori} moral concepts agreed by the entire network. Notice that neither choice is directly associated to objective physical or psychological harm, but is based on the subjective acceptance of this actions in the agent's network (group, community, population, culture etc). 

The action of $i$ on $j$ is then represented by the matrix element $J_{ij}\in\chs{0,\pm1}$, where -1 indicates a  \emph{harmful} action and +1 a \emph{helpful} one. The zero value corresponds to no action at all. In the present model, we will restrict this value only to the diagonal entries of the matrix $J$, i.e., those of the form $J_{ii}$ which represent self-interactions. The matrix $J$ \emph{is not symmetric} in general as the order of indexes \emph{is} important, the first representing the agent that is originating the action and the second the one being targeted by it as depicted in fig.\ref{figure:FCN} (right).  

To each individual or node of the network, we associate a \textit{personality vector} $\pi_i=(u_i,w_i)$ with binary components $u_i,w_i\in\chs{\pm1}$. These two components represent ``emotional desires'' of the individual $i$. The first component, $u_i$, indicates how the agent $i$ would ``desire'' to act towards another agent. When $u_i=-1$, this indicates that agent $i$ prefers to do harmful actions, while a value +1 would indicate a preference for helpful ones. The assumed \emph{emotional} nature of these preferences is intended to mean that fulfilling them leads to a subconscious satisfaction of the agent, which one can roughly associate to human feelings. A sadistic personality, for instance, would be represented by $u_i=-1$ as the agent feels pleasure by harming others, while $u_i=+1$ would represent an altruistic agent who enjoys helping its peers.

The second component of the personality vector represents how the agent would like to be treated by others. A $w_i=+1$ is what one would expect from someone who would appreciate the help of others, a behavior probably seen as ``normal'', while $w_i=-1$ corresponds to an agent that prefers to be mistreated, as would be the case of a ``masochist'' agent, for instance.

There are clearly several simplifications in the above personality attributions compared to real situations. First, we consider that both $u_i$ and $w_i$ depend only on the agent $i$. They are the same independently of which agent is the target of the action. A more sophisticated version of this model would consider that this preference might vary according to the target and the agent's feelings towards it. For instance, $i$ might be emotionally led to act differently whether it is interacting with its own mother or with a well-known serial killer.

Another simplification, which is less important for the situation we are going to study, concerns the fact that the personality vector might change with time. Individual emotional responses are not only genetically defined, but are affected by interaction with the environment. The scenario studied in this work does not deal with dynamics and, therefore, this simplification is not relevant in the present case. It is however improbable that a significant change in behavior that can affect the whole population occurs in a short time scale (although it is not impossible to happen) and, therefore, we believe that we can consider stable personalities for a certain macroscopic period of time as a first approximation.

The use of binary vectors to characterize the agents' personalities represents also a strong simplification. Human behavior has a wide spectrum of variability which would be better modeled by continuous rather than binary variables. The intention of the present model, however, is to take a first step towards this modeling and try to identify some general basic principles and emergent behaviors. It is well-known that simplified models are used even in social sciences and, in particular, in psychology, with the best known example being the Myers-Brigg Type Indicator \cite{Quenk09} which classifies human behavior by considering only 14 types and is largely used by institutions to actually select prospective employees. 

Given the two \textit{personality components} of $\pi_i$, whether or not agent $i$ feels fulfilled by the interactions within its network will be represented by its \emph{satisfaction}
\begin{equation}
  \sigma_i = \sgn\col{\Omega(\pi,J,\gamma)},
\end{equation}  
with
\begin{equation}
  \Omega(\pi,J,\gamma) = \frac1{N} \col{\gamma U_i+(1-\gamma)W_i},
  \label{equation:Omega}
\end{equation}
and
\begin{equation}
  U_i = u_i  \sum_{j\neq i} J_{ij}, \qquad  W_i = w_i \sum_{j\neq i} J_{ji}.
\end{equation}

This definition allows three values for the agent's satisfaction. When $\sigma_i=+1$, we say that the agent feels satisfied, when $\sigma_i=-1$ it feels dissatisfied and when $\sigma_i=0$ the agent is neither.  

The scaling $1/N$ in $\Omega$ is used to make it an intensive quantity in the number of agents, keeping it finite when $N\to\infty$. This is the limit in which we are interested as it is only when the number of agents is \emph{exactly} infinite that we can unambiguously identify the model's phase transitions.  

The extensive quantity $U_i$ is the sum of all contributions to the fulfillment of the agent's desire on how to treat other agents, while the also extensive $W_i$ represents the same concerning how the agent feels treated by the others. The parameter $\gamma$ is taken from the interval $\col{0,1}$ for convenience and represents the relative \emph{emotional} importance given by the individual to each of these terms. This parameter will be kept fixed and will be the same for the whole network.

We recall that in the static version analyzed here, each realization of the whole experiment will be considered as consisting of a fixed configuration of $J$ and $\pi$ which however varies from one realization to another. Random disorder enters the model through the distributions of possible values of these two variables in each realization of the ``experiment'' which might involve a totally different set of individuals and interactions each time. 

Notice that the satisfaction contains only ``emotional'' contributions. Human beings have mechanisms to suppress emotional responses under rational considerations. Therefore, a term based on each agent's level of rationality can be devised which would compete with its emotional satisfaction. However, such a consideration is out of the scope of the present paper and should be addressed in more refined versions of the model. 

For the present calculation, we assume that the personality parameters are i.i.d. with distribution
\begin{align}
  \prob{u} &= (1-s)\delta\prs{u,1}+s\delta\prs{u,-1}, \\
  \prob{w} &= (1-m)\delta\prs{w,1}+m\delta\prs{w,-1},
\end{align}
and $s,m\in\col{0,1}$ ($s$ and $m$ the same for all agents).

A natural Hamiltonian for the whole network would be the negative of its net satisfaction $H=-\sum_i \sigma_i$ with the distribution of personalities being the quenched disorder and the moral decisions $J_{ij}$ the dynamical variables (analogous to the spins in a magnetic model). This would lead to a thermodynamic distribution for $J$ at inverse temperature $\beta$ given by $\prob{J|\pi} \propto e^{-\beta H}$. Such an equilibrium distribution corresponds to a dynamics in which agents are driven to act towards the minimization of the network's energy, a behavior that can be classified as \emph{rational}. As discussed in the Introduction, there are several reasons why agents might not act rationally. For instance, they might lack either the will or the resources to follow the strategy based on minimizing the energy. This is actually the case for our model and, therefore, the distribution of decisions will not be the equilibrium one for this natural choice of Hamiltonian. 

In the particular case in which every agent has the same $\gamma$, the strategies that minimize the Hamiltonian are very simple and easy to see. If $\gamma>1/2$, then more importance is given for $U_i$ than for $W_i$ and it benefits a totally selfish strategy. The opposite is true if $\gamma<1/2$. These two strategies, however might not be optimal in general outside these parameter ranges. In fact, we will show that there are phase transitions in the average satisfaction of the networking (to be defined in the following) when using these strategies outside their optimality zone.  

The specific distribution of the agent's actions $J_{ij}$ we are going to use is, in fact, a key feature of this work. In addition to the thermal equilibrium distribution and the two simplified strategies considered above, there is clearly an infinite number of ways for an agent to choose how to act towards another one. For instance, all agents could simply act in the same way by choosing $J_{ij}=+1$ independently of $\pi_i$ or $\pi_j$. In order to be able to relate our work to empirical facts, we focus on what we described in the introduction as $n$-th order moral decisions based on Piaget's Ladder, which result in the values for the matrix $J$ given in table \ref{table:nom}.

\begin{table}
  \centering
  \begin{tabular}{|c|c|c|}
    \hline
    $J_{ij}$        & Order & Moral Behavior \\
    \hline
    $u_i$           & 0     & Selfish        \\
    $w_i$           & 1     & Parental       \\
    $w_j$           & 2     & Empathetic     \\
    \hline       
  \end{tabular}
  \caption{$n$-th order morals}
  \label{table:nom}
\end{table}

\begin{figure}
  \centering
  \includegraphics[width=6cm]{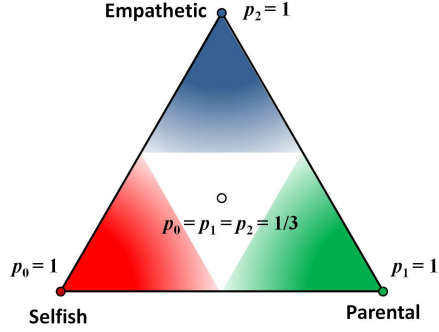}
  \caption{Graphical representation of moral strategies as points of a triangle.}
  \label{figure:ET}
\end{figure}

Here we consider the distribution of moral decisions defining the action of agent $i$ towards agent $j$ to be given by a convex combination of the $n$-order moral decisions as
\begin{equation}
  \prob{J_{ij}|\pi_i, \pi_j} = p_0\delta\prs{J_{ij},u_i}+p_1\delta\prs{J_{ij},w_i}+p_2\delta\prs{J_{ij},w_j},
  \label{equation:action}
\end{equation}
with
\begin{equation}
  \sum_i p_i = 1,
\end{equation}
which can be represented as points in an equilateral triangle as in fig. \ref{figure:ET}.

The probabilities in the above equation can be seen either as the proportion of individuals choosing one of the three levels of moral decision or as the probability of one individual taking each of them at a certain time. Because the 2nd order decision requires knowledge of the personality vector of others, we will assume that agents possess full noiseless information about the $w$ component of all other agents. A more realistic scenario would be when only partial information is available, adding an extra source of disorder.

The \emph{average satisfaction} of the whole network can then be measured by averaging the individual satisfactions over the decision strategy and over the disorder in the personality vectors as
\begin{equation}
  S = \avg{\sigma_k}{\vc{u}, \vc{w},J},
\end{equation}
in which the index $k$ inside the average is only for convenience as it disappears due to the averaging over all variables $u_i$, $w_i$ and $J_{ij}$ represented by the vectors $\vc{u}$ and $\vc{w}$ and the matrix $J$. This is equivalent to averaging over frozen realizations of the system. 

We will take $S$ as the order parameter defining the phases of the model. Ordered phases correspond to situations where $S=+1$, with all agents but for a set of measure zero are satisfied, and $S=-1$, when all agents are dissatisfied. Intermediate values of $S$ correspond to disordered phases. In the next section, we analyze the different behavior of the system as the parameters of the model are varied. 

\section{Phase Structure}
\label{section:PS}

In the limit $N\to\infty$ of an infinitely sized network, one can obtain the exact value of the average satisfaction for general values of the strategy's probabilities as (see appendix \ref{appendix:AS} for details)
\begin{equation}
  S = \avg{\sgn\,\mu}{u,w},
\end{equation}
where
\begin{equation}
  \mu = p_0\col{\gamma+(1-\gamma)\bu w}+p_1\col{\gamma u w+(1-\gamma)\bw w}+p_2\col{\gamma \bw u+(1-\gamma)},
\end{equation}
and
\begin{equation}
  \bu = \avg{u}{}=(1-2s), \qquad \bw = \avg{w}{}=(1-2m).
\end{equation}

In the following, we will analyze the four special cases represented by the points in fig. \ref{figure:ET}.   

\subsection{0th Order Moral}

When $p_0=1$ then $p_1=p_2=0$ and we are assuming that all agents are acting selfishly, or taking 0th order moral decisions. Agents act by ignoring other agents' desires, maximally satisfying their own $U_i$ in every decision. The expression for the average satisfaction simplifies to
\begin{equation}
  S = (1-m)\,\sgn\col{\gamma+(1-\gamma)(1-2s)}+m\,\sgn\col{\gamma-(1-\gamma)(1-2s)}.
\end{equation}

This confirms the result we have discussed before that, when $\gamma>1/2$, the system is in the ordered phase $S=+1$ for all values of $m$ and $s$ as each agent's satisfaction depends more on how it acts on other agents than on how the others act on it.  

The case $\gamma=1/2$ is degenerate, giving
\begin{equation}
S = (1-m)\,\sgn\,(1-s)+m\,\sgn\,s.
\end{equation} 

As $s\in\col{0,1}$, the above expression gives 1 for all values of $m$ except for $s=0,1$, in which case linear relations with $m$ are obtained. Because these are simply one dimensional lines on the borders of the diagram, they cannot be easily seen. This seems somewhat puzzling as completely selfish moral decisions from every single person are guaranteeing the well-being of the whole network even when both terms in the satisfaction have equal weights. The balance between the two competing terms $U_i$ and $W_i$ for $\gamma=1/2$ is however very delicate. The strategy $p_0=1$ guarantees that everyone has always the maximum value for its $U_i$, which means $U_i/N\to1$ in the thermodynamic limit. This implies that any small deviation from complete dissatisfaction concerning the way agents are being treated will top the balance to the side of a satisfied population.

\begin{figure}
  \centering
  \includegraphics[width=16cm]{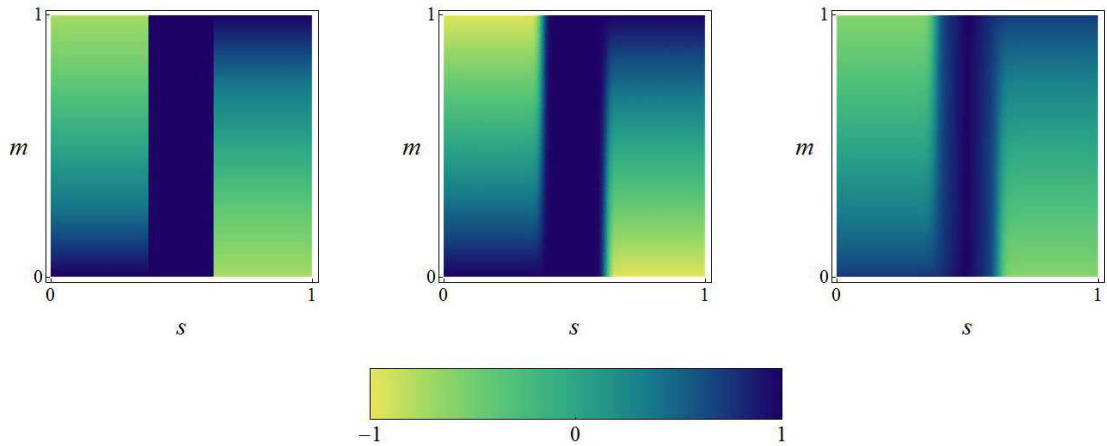}
  \caption{Phase diagrams of selfish decisions for $\gamma=0.2$ (color online). Left: exact result for the infinite system; middle: computer simulation with $N=1000$ averaged over 40 independently generated configurations; right: leading order analytical approximation for $N=1000$. The central band (online red) is totally ordered while the order parameter varies linearly from -1 to 1 in the lateral bands from top to bottom in the left band and from bottom to top in the right band.}
  \label{figure:p0_01}
\end{figure}

The most interesting cases are when $\gamma<1/2$, which means that more importance is given to the way the agent is treated by others than to the way it treats other agents. Fig. \ref{figure:p0_01} shows the result of the exact expressions and simulations for $\gamma=0.2$. The simulations are run with a population size of $N=1000$ agents and averaged over 40 independently generated realizations of $J$ and $\pi$ configurations. The diagram at the center is the result of the simulation, the one at the left is the theoretical value for $N=\infty$. The difference between the two diagrams is due to finite size effects. 

The diagram to the right represents the leading order theoretical correction for the system's finite size. Because $\sigma^2$ in equation (\ref{equation:fs}) from appendix \ref{appendix:AS} is identically zero when $p_0=0$, we need to consider the variance coming from the average over the $u_i$'s. This leads to the expression
\begin{equation}
  S(N) = (1-m)\,\erf\col{\frac{\gamma+(1-\gamma)(1-2s)}{\sqrt{2\col{1-(1-2s)^2}/N}}}+
         m\,\erf\col{\frac{\gamma-(1-\gamma)(1-2s)}{\sqrt{2\col{1-(1-2s)^2}/N}}}.
\end{equation}
 
One can see that a rich phase structure in the diagram develops with the appearance of three very distinctive bands. The central band represents the ordered phase $S=1$ and its width decreases with $\gamma$. This width can be calculated in the following way. Let us assume that $\gamma=1/2-\epsilon$ where $0<\epsilon\leq1/2$. The ordered band requires the arguments of the two sign functions to be positive independently of the value of $m$, i.e.
\begin{equation}
  \gamma+(1-\gamma)(1-2s)>0, \qquad \gamma-(1-\gamma)(1-2s)>0.
\end{equation}

This implies
\begin{equation}
  \frac{2\epsilon}{1+2\epsilon}<s<\frac{1}{1+2\epsilon},
\end{equation}
and the bandwidth is therefore
\begin{equation}
  \delta = \frac{1-2\epsilon}{1+2\epsilon}.
\end{equation}

Notice that the area in which the whole network is fully satisfied with the distribution of moral decisions is larger than or \emph{at most} equal to that in which it is not. This holds for \emph{any} value of $\gamma$ due to the symmetry of the lateral bands. This is a disheartening result as it suggests that trying to derive moral decisions from rational principles that benefit society as a whole in an objective way might not be attainable. Such a principle would force one to accept that completely selfish decisions are morally correct, even when they are meant to do harm to those who do not want it. Of course this result is based on a simplified model, but it clearly illustrates that tying moral concepts to some rational measure of social well-being does not work in general.

\begin{figure}
  \centering
  \includegraphics[width=10cm]{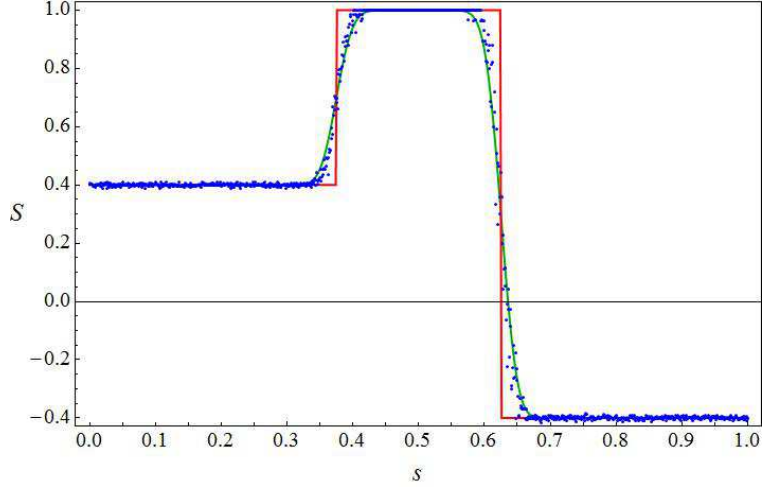}
  \caption{The graph shows the order parameter $S$ along a line of constant $m=0.3$ in the phase diagrams of fig. \ref{figure:p0_01} (color online). The dots (online blue) represent the values of the simulations, the smooth line closer to them (online green) is the leading order finite size approximation and the non-smooth distribution (online red) is the infinite system. Notice that in this case $\gamma=0.2$, giving a width of 0.25 for the ordered central band with edges at $s=0.375$ and $s=0.625$.}
  \label{figure:p0_PT}
\end{figure}

The diagram is also interesting from the statistical physics point of view. By keeping $m$ constant and varying $s$ from zero to one, we pass through two discontinuous phase transitions. We start with a disordered phase which becomes ordered at the central band and then becomes disordered once again, but with an average satisfaction with the opposite sign. These transitions are shown in fig. \ref{figure:p0_PT}. Inside each of the two disordered bands, by keeping $s$ fixed and varying $m$ in the interval $[0,1]$, we have a continuous linear crossover between the two ordered phases with opposite signs of satisfaction. 

\begin{figure}
  \centering
  \includegraphics[width=10cm]{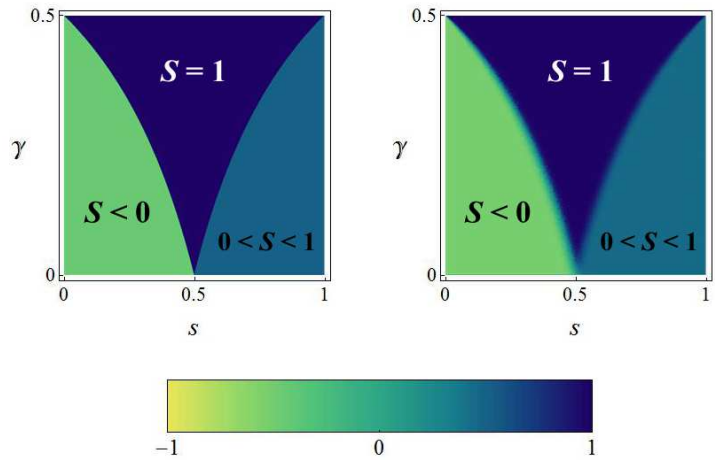}
  \caption{Phase diagram (color online) showing the discotinuous transitions when $\gamma$ and $s$ are varied for constant $m=0.8$. The left graph shows the result for an infinite system, while the right one shows the simulated diagram for $N=1000$.}
  \label{figure:QPT0}
\end{figure}

Another interesting phase diagram is obtained by plotting the values of $S$ in a $\gamma\times s$ diagram, what is shown if fig. \ref{figure:QPT0} for the fixed value $m=0.8$. One can clearly see the quick growth of the central ordered band as the parameter $\gamma$ increases from zero to 1/2.

\begin{figure}
  \centering
  \includegraphics[width=10cm]{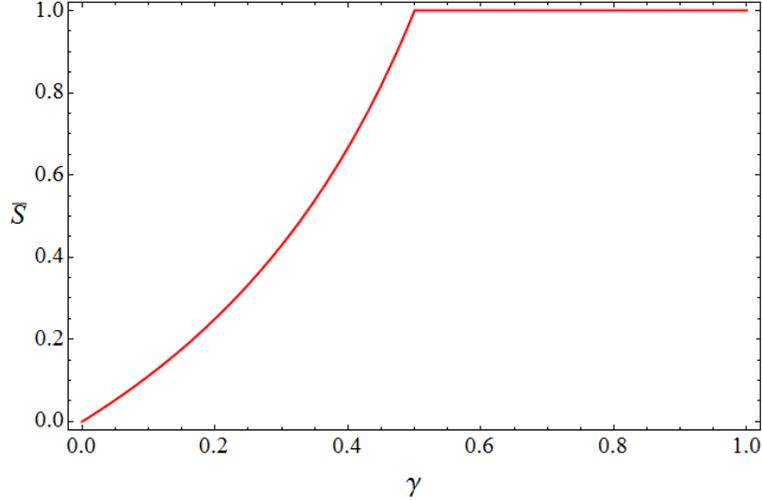}
  \caption{Continuous phase transition from a disordered to an ordered phase at $\gamma=1/2$ with the order parameter taken as the average over the disorder in the personality vectors of the network satisfaction ($\bar{S}$) for 0th order moral decisions.}
  \label{figure:bandsize}
\end{figure}

Fig. \ref{figure:bandsize} shows the continuous phase transition resulting from the change in the ordered band width at $\gamma=1/2$ by taking as the order parameter $\bar{S}$, the average of $S$ over the disorder parameters $s$ and $m$.  

\subsection{2nd Order Moral}

From the moral point of view, one would be more inclined to accept that 2nd order moral decisions are better, or more considerate, choices than 0th order ones. The former is based on the idea of respecting others, while the latter bears no consideration for other's feelings. By construction, the model we are using has a symmetry connecting these two moral strategies. The substitution
\begin{equation}
  \gamma \to 1-\gamma, \qquad u_i \to w_i, \qquad J_{ij} \to J_{ji},
\end{equation}
keeps all $\sigma_i$ the same and, therefore, also the average satisfaction $S$. The effect on $S$ is equivalent if we change the last substitution by
\begin{equation}
  p_0=1 \to p_2=1.
\end{equation}

Therefore, if we change from selfish to empathetic moral decisions, the diagrams exchange their vertical and horizontal axes at the same time as $\gamma\to 1-\gamma$. As in the selfish case, the area of the diagram in which the satisfaction is positive is always larger than the negative one for all values of $\gamma$. If it was not for the fact that the same happens with the selfish behavior, that would be good news.

Although our model is quite simple, it is based on reasonable enough assumptions. Given the agreed concepts of morality, one can then see that two completely opposite moral decisions taken by the \emph{whole} population guarantee the well-being of the network in the majority of the parameter space. This implies that, if one bases the concept of morality on ``rational'' arguments concerning the overall well-being, both behaviors should be considered morally right. None is more harmful to the society than the other.

This result seems paradoxical, but it arises from associating moral to satisfying the majority. Satisfying the majority is many times equivalent to ignoring or oppressing the minority, which is morally not acceptable. The paradox disappears if one recognizes that what can be associated with the definition of morally acceptable behavior is in fact the value of $\gamma$. We tend to consider behaviors to be morally acceptable when everyone is being respected by others, which is equivalent to count only the term $W_i$ for the satisfaction. In other words, this is the situation when $\gamma=0$. For the selfish strategy, a trivial calculation shoes that this results in $\bar{S}=0$ while for the empathic strategy this gives $\bar{S}=1$, the expected result for a morally accepted behavior.     

There are many repressive moral beliefs in all societies which vary with time and that are not agreed by all individuals. For instance, the well-known repression of homosexual individuals in World War II in Britain, which was the probable cause of Alan Turin's suicide, was considered a correct moral behavior at that time. In India and many other parts of the world, arranged marriages are still in practice even though the bride might not have a say in the final decision. Several religions have strict dressing codes and enforcing it, even violently in some cases, are considered as moral behavior even when it goes against the wishes of the individuals. However, these behaviors would all fall in the $\bar{S}=0$ region for $\gamma=0$, which is a culturally independent criteria.  

\subsection{1st Order Moral}

There is no symmetry connecting 0th and 2nd order moral decisions to 1st order ones. Taking only 1st order moral decisions creates a different phase diagram. The argument of the sign function simplifies to
\begin{equation}
  \mu = \gamma uw+(1-\gamma)(1-2m)w,
\end{equation}
and the average satisfaction is then
\begin{equation}
  S = (1-2m)\chs{(1-s)\,\sgn\col{\gamma+(1-\gamma)(1-2m)}+s\,\sgn\col{-\gamma+(1-\gamma)(1-2m)}}.
\end{equation}

\begin{figure}
  \centering
  \includegraphics[width=11cm]{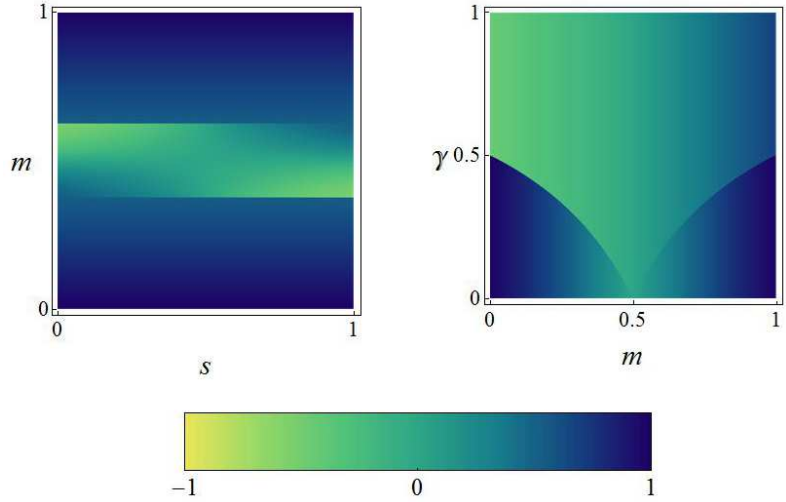}
  \caption{Phase diagrams for the 1st order moral decisions (color online). Left: $s\times m$ phase diagram for $\gamma=0.2$ showing the discontinuous transitions on the borders of the central band. Right: $m\times\gamma$ diagram for $s=0.8$. One can see the continuous transition in which the central band disappears completely at the value $\gamma=1/2$.}
  \label{figure:p1pds}
\end{figure}

\begin{figure}
  \centering
  \includegraphics[width=10cm]{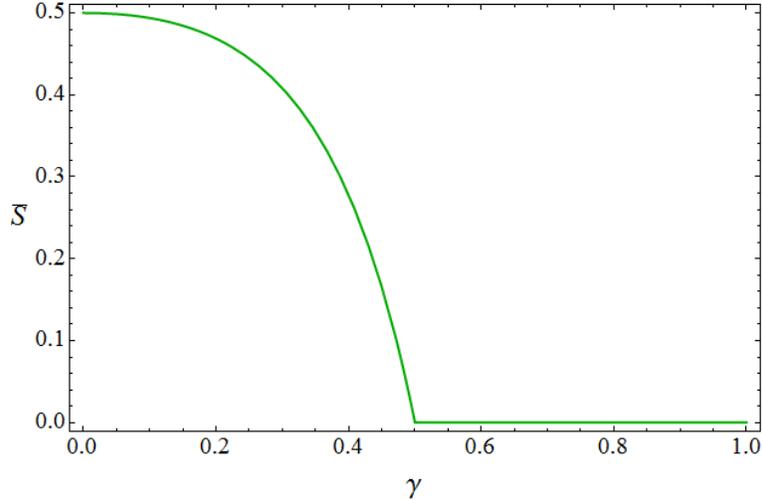}
  \caption{Continuous phase transition from a partially ordered ($\bar{S}=1/2$) to a disordered phase at $\gamma=1/2$.}
  \label{figure:p1pt}
\end{figure}

Analogous phase diagrams to the ones for the other strategies are given in fig. \ref{figure:p1pds} which shows a very similar structure, but with different details of the phases and transitions. For instance, the $s\times m$ diagram still presents a central band, which is however not totally ordered. It is also of opposite sign compared to the 0th and 2nd order strategies, with the central band indicating a disordered phase with negative satisfaction. The limits and size of the band are however the same as for the other strategies, which also implies the presence of a continuous phase transition in $\bar{S}$ when $\gamma$ is varied. This transition, depicted in fig. \ref{figure:p1pt}, is however between a partially ordered and a totally disordered phase, differently from those for the other strategies.

This also shows that when $\gamma=0$ we have $\bar{S}=1/2$. If we use the criteria suggested before, this is still a behavior which should not be considered morally acceptable, but would be \emph{more} acceptable than the completely selfish decisions.

\subsection{Mixed Strategy}

\begin{figure}
  \centering
  \includegraphics[width=12cm]{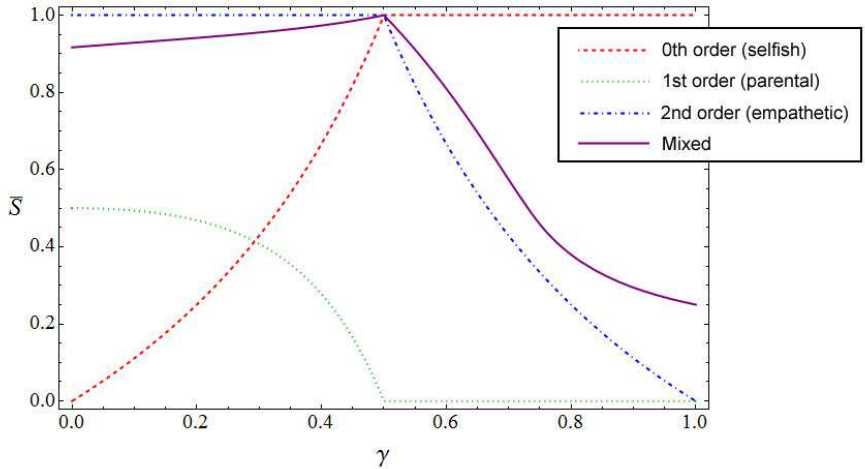}
  \caption{Continuous phase transitions on $\bar{S}=1/2$ for the four strategies at $\gamma=1/2$.}
  \label{figure:mixed}
\end{figure}

For the sake of completeness, we will briefly consider a mixed strategy for which $p_0=p_1=p_2=1/3$. This strategy is unlikely to appear in real life as it has no well-defined rational behind it. Fig. \ref{figure:mixed} shows a comparison between the $n$-order moral decisions and the mixed strategy. All strategies present continuous phase transitions in $\bar{S}$ for $\gamma=1/2$ as is evident from the discontinuities in the derivatives at that point. Although the mixed strategy remains as the second best throughout all range of $\gamma$, it only attains $\bar{S}=1$ at the critical point.

\begin{figure}
  \centering
  \includegraphics[width=10cm]{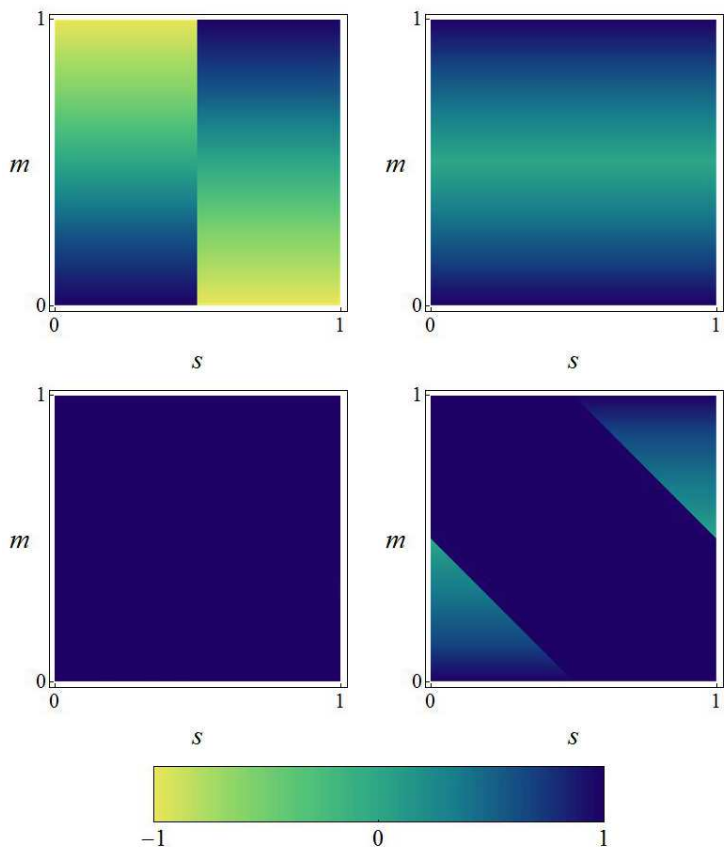}
  \caption{Phase diagrams for the four strategies at $\gamma=0$ (color online). From left to right. Top row: 0th order, 1st order. Bottom row: 2nd order, mixed.}
  \label{figure:allpds}
\end{figure}

As discussed before, it seems reasonable to associate some sort of \emph{degree of morality} to each strategy by their values at $\gamma=0$. It is then interesting to note that by randomly choosing each strategy with the same probability each time leads to a much higher moral degree than the 0th and 1st order strategies. One might be tempted to argue that some respect once in a while is much better than none. This striking difference can be better understood by looking at the phase diagrams $s\times m$ for all strategies at $\gamma=0$ given in fig. \ref{figure:allpds}. 

\section{Conclusions}
\label{section:Conc}

In this work we analyzed a model of agents interacting in a fully connected network by taking moral decisions. The agent's decisions were constrained by an increased level of empathy based on the stages of Piaget's theory of cognitive development. Within this model, we defined a measure of emotional satisfaction related to the fulfillment of the agents' personal desires to characterize the phases of the network as the parameters of the model are varied. These desires were modeled by local random binary fields at each site of the network and were called \emph{personality vectors} as they represent personality traits of the agents. 
 
The introduced model is exactly solvable. We were able to calculate analytically the value of the average satisfaction of the network in the limit of an infinite number of agents and also its leading order contribution in powers of $1/N$. Using the average satisfaction as an order parameter, we then were able to find the model's phase diagrams and to identify the existence of both continuous and discontinuous transitions triggered by the variation of the disorder parameters.
 
We chose to analyze four different specific strategies of moral decisions in this work. In the first three, agents take decisions according to each stage in Piaget's theory, a three-steps hierarchy to which we gave the name of \emph{Piaget's Ladder}. Each step in this ladder is associated with an $n$-th order moral. The 0th order corresponds to completely selfish decisions and the 2nd order to completely empathetic ones. The 1st order corresponds to an intermediate case. In addition to these three strategies, we also analyzed the case in which each agent takes a decision by randomly choosing one of those strategies at each time. 

Interestingly, there is a symmetry between the 0th and 2nd order moral decisions resulting int he fact that the selfish strategy guarantees the \emph{average} well-being of the network as well as the empathetic one. Although the model is very simple, its assumptions are reasonable enough to indicate that moral beliefs cannot be simply based on rational minimization of some energy function as two strategies that would clearly be considered morally opposite by most people lead to the same phase diagrams. One possibility is that the parameters of the disorder are tuned in humans to values for which a species-wide agreement on moral concepts can be derived. One solution to this conundrum suggested by us was that a moral degree might be associated not to the overall satisfaction, but to its particular case in which the satisfaction of others is more important than ours. This leads to a sensible classification when all four strategies are compared. Of course, a more realistic view, as that of MFT, would require a more sophisticated set of parameters, but this is out of the scope of the present analysis.

The 0th and 2nd order strategies minimize the Hamiltonian in the appropriate domains of the parameter $\gamma$ which measures the relative importance of the two terms in the satisfaction. As expected, the 1st order and the mixed strategies underperform those strategies in their optimal ranges, although the mixed strategy seems to be the one which remains more acceptable in a wider range of $\gamma$.

From the technical point of view, the present model has many interesting properties. It is simple enough to be easily interpretable and completely solvable. At the same time, it presents a range of very interesting phase diagrams and transitions. We have discussed only some basic features of these transitions due to size constraints of this paper, choosing to focus on the interpretation of these results. In a forthcoming paper, we intend to explore the interesting mathematical structure of this model in more detail.   
 
Finally, there are many directions in which this model can be extended in order to include more realistic behavior. For instance, different network topologies can be used. One could also devise a scenario in which agents try to infer by a learning algorithm the desires of others. This would bring dynamics into the model and require a more sophisticated treatment which we will leave for a future study. The personality vector also can be extended either to a higher dimensionality to include more realistic personality variations or to continuous instead of binary values. We intend to explore these extensions in future works.    

\section*{Acknowledgments}

I would like to thank Dr Juan Neirotti for very useful discussions and comments.


\bibliographystyle{apsrev4-1}
\bibliography{moral}

\appendix

\section{Analytical Calculation of the Average Satisfaction}
\label{appendix:AS}

We can explicitly write the expression for the average satisfaction $S$ as
\begin{equation}
  \begin{split}
    S = \avg{\sigma_k}{\vc{u}, \vc{w},J} &= \sum_{\chs{u_i}}\sum_{\chs{w_i}}
                                                   \sum_{\chs{J_{ij}}}\col{\prod_i \prob{u_i}\prob{w_i}}
                                                   \col{\prod_{\substack{i,j \\ i\neq j}} 
                                                   \prob{J_{ij}|u_i,w_i,u_j,w_j}}\\
                                                &  \quad\times
                                                   \sgn\col{\frac1N\prs{\gamma u_k\sum_{l\neq k}J_{kl}
                                                   +(1-\gamma)w_k\sum_{l\neq k}J_{lk}}}.                                           
  \end{split}
\end{equation}

We start by doing the average over $J$. The argument of the sign contains averages over $N$ variables $J_{ij}$. Although they are not identically distributed, the fact that they are independent given $\vc{u}$ and $\vc{w}$ allows us to use an extension of the Central Limit Theorem (CLT). In the following, we explicitly present this extension. 

By means of a Dirac delta distribution, one can write the above equation as
\begin{equation}
  S = \avg{\int \frac{dx\,d\hat{x}}{2\pi}\; e^{ix\hat{x}} \prs{\mbox{sgn}\,x} \Gamma(\hat{x},\vc{u},\vc{w},\gamma)}
      {\vc{u},\vc{w}},
\end{equation}
with
\begin{equation}
  \Gamma = \sum_{\chs{J_{ij}}}\col{\prod_{\substack{i,j \\ i\neq j}}\prob{J_{ij}|u_i,w_i,u_j,w_j}}
           \exp\chs{-\frac{i\hat{x}}N\col{\gamma u_k\sum_{l\neq k}J_{kl}+(1-\gamma)w_k\sum_{l\neq k}J_{lk}}}.
\end{equation}

The average can now be factorized and written as
\begin{equation}
  \prod_{l\neq k} \Lambda^1_{lk} \Lambda^2_{lk}=\exp\chs{\sum_{l\neq k} \prs{\ln \Lambda^1_{lk}+\ln \Lambda^2_{lk}}},
\end{equation}
where
\begin{align}
  \Lambda^1_{lk} &\equiv \sum_{J_{kl}} \prob{J_{kl}} \exp\chs{-\frac{i\hat{x}}N \gamma u_k J_{kl}},\\  
  \Lambda^2_{lk} &\equiv \sum_{J_{lk}} \prob{J_{lk}} \exp\chs{-\frac{i\hat{x}}N (1-\gamma)w_k J_{lk}},                  
\end{align}
with $k$ a fixed index.

Given that $J$ is a binary matrix, we can rewrite the probability distributions as
\begin{equation}
  \prob{J_{ij}|\pi_i, \pi_j} = p_0\prs{\frac{1+J_{ij}u_i}2}+p_1\prs{\frac{1+J_{ij}w_i}2}+p_2\prs{\frac{1+J_{ij}w_j}2},
\end{equation}
which gives
\begin{align}
  \Lambda^1_{lk} &= \cos\col{\frac{\gamma\hat{x}}N}-i\prs{p_0+u_k w_k p_1+u_k w_l p_2}
                    \sin\col{\frac{\gamma\hat{x}}N}\\
  \Lambda^2_{lk} &= \cos\col{\frac{(1-\gamma)\hat{x}}N}-i\prs{u_l w_k p_0+w_l w_k p_1+ p_2}
                    \sin\col{\frac{(1-\gamma)\hat{x}}N}.
\end{align}

Expanding the cosines, sines and logarithms up to order $1/N^2$, we get
\begin{align}
  \ln \Lambda^1_{lk} &\approx -\frac{\gamma^2\hat{x}^2}{2N^2}\col{1-\prs{\lambda^1_{kl}}^2}
                              -i\frac{\gamma\hat{x}}N \lambda^1_{kl},\\ 
  \ln \Lambda^2_{lk} &\approx -\frac{(1-\gamma)^2\hat{x}^2}{2N^2}\col{1-\prs{\lambda^2_{kl}}^2}
                              -i\frac{(1-\gamma)\hat{x}}N \lambda^2_{kl},                                
\end{align}
where
\begin{align}
  \lambda^1_{kl} &= p_0+u_k w_k p_1+u_k w_l p_2,\\
  \lambda^2_{kl} &= u_l w_k p_0+w_l w_k p_1+ p_2.  
\end{align}

By defining
\begin{align}
  \mu      &\equiv \frac1N\sum_{l\neq k}\col{\gamma\lambda^1_{kl}+(1-\gamma)\lambda^2_{kl}},\\ 
  \sigma^2 &\equiv \frac1{N^2}\sum_{l\neq k}\chs{\gamma^2\col{1-\prs{\lambda^1_{kl}}^2}
                   +(1-\gamma)^2\col{1-\prs{\lambda^2_{kl}}^2}},
\end{align}
we can write 
\begin{equation}
  \begin{split}
    S &= \avg{\int \frac{dx\,d\hat{x}}{2\pi}\;e^{-\frac{\hat{x}^2\sigma^2}{2}+i\hat{x}(x-\mu)} \prs{\mbox{sgn}\,x}}
         {\vc{u},\vc{w}}\\
      &= \avg{\mbox{erf}\prs{\frac{\mu}{\sqrt{2\sigma^2}}}}{\vc{u},\vc{w}}.   
  \end{split}
\end{equation}

Notice that this is the extension of the CLT that we alluded to. The average over $J$ became an average over a Gaussian distributed variable $x$ with mean $\mu$ and variance $\sigma^2$ which are what we would obtain by calculating the mean and variance of each $J_{ij}$ and doing the appropriate linear combination or, using the notation of equation (\ref{equation:Omega}), 
\begin{align}
  \mu      &= \frac1N\sum_{l\neq k}\col{\gamma u_k\avg{J_{kl}}{}+(1-\gamma)w_k \avg{J_{lk}}{}},\\
  \sigma^2 &= \frac1{N^2}\sum_{l\neq k}\col{\gamma^2 (1-\avg{J_{kl}}{}^2)
              +(1-\gamma)^2(1-\avg{J_{lk}}{}^2)},
\end{align}
where
\begin{equation}
  \avg{J_{ij}}{} = p_0 u_i + p_1 w_i + p_2 w_j.
\end{equation}

The expressions for $\mu$ and $\sigma^2$ can be further simplified for $N\to\infty$
\begin{equation}
   \begin{split}
    \mu &= \frac1N \sum_{l\neq k}\col{\gamma\prs{p_0+u_k w_k p_1+u_k w_l p_2}+(1-\gamma)
           \prs{u_l w_k p_0+w_l w_k p_1+ p_2}}\\
        &= p_0\col{\gamma+(1-\gamma)\bu w_k}+p_1\col{\gamma u_k w_k+(1-\gamma)\bw w_k}
           +p_2\col{\gamma \bw u_k+(1-\gamma)}, 
   \end{split}
 \end{equation}
and
\begin{equation}
  \begin{split}
   \sigma^2 &= \frac1{N^2} 
               \sum_{l\neq k}\chs{\gamma^2\col{1-(p_0+u_k w_k p_1+u_k w_l p_2)^2}
               +(1-\gamma)^2\col{1-(u_l w_k p_0+w_l w_k p_1+ p_2)^2}}\\
            &= \frac1N\left\{\col{\gamma^2+(1-\gamma)^2}(1-p_0^2-p_1^2-p_2^2)
               -2p_0 p_1\col{\gamma^2 u_k w_k +(1-\gamma)^2\bC}\right.\\
            &  \quad\left.
               -2p_0 p_2\col{\gamma^2 u_k\bw +(1-\gamma)^2 \bu w_k}
               -2p_1 p_2\col{\gamma^2 +(1-\gamma)^2}w_k\bw\right\}.
   \end{split}
   \label{equation:fs}
\end{equation}
where we introduced the definitions
\begin{equation}
  \bu = \frac1N\sum_{l\neq k}u_l, \qquad \bw = \frac1N\sum_{l\neq k}w_l, \qquad
  \bC = \frac1N\sum_{l\neq k}u_l w_l.
\end{equation}

The CLT can now be directly applied in its original form to the above variables, resulting in 
\begin{equation}
  S = \avg{\mbox{erf}\prs{\frac{\mu}{\sqrt{2\sigma^2}}}}{u_k,w_k,\bu,\bw,\bC},   
\end{equation}
where the hatted variables are distributed according to the following Gaussian distributions
\begin{align}
  \bu &\sim \cN\prs{\bu|\avg{u_i}{},\sigma^2_u}, \\
  \bw &\sim \cN\prs{\bw|\avg{w_i}{},\sigma^2_w}, \\
  \bC &\sim \cN\prs{\bu|\avg{u_i}{}\avg{w_i},\sigma^2_{uw}}, 
\end{align}
where
\begin{equation}
  \cN\prs{y|\mu_y,\sigma^2_y} \equiv \frac{e^{-\frac{(y-\mu_y)^2}{2\sigma^2_y}}}{\sqrt{2\pi\sigma^2_y}},
\end{equation}
and
\begin{align}
  \avg{u_i}{}&=(1-2s), \qquad \sigma^2_u=1- \avg{u_i}{}^2=4s(1-s),\\
  \avg{w_i}{}&=(1-2m), \qquad \sigma^2_w=1- \avg{w_i}{}^2=4m(1-m),\\
  \sigma^2_{u_i w_i} &=1-\avg{u_i}{}^2\avg{w_i}{}^2. 
\end{align}

We do not need to carry the index $k$ anymore and therefore we write $u$ and $w$ instead of $u_k$ and $w_k$. In the limit $N\to\infty$, the above Gaussians become Dirac deltas in their means and the error function becomes the sign of its argument, which gives our final expression
\begin{equation}
  S = \avg{\sgn\,\mu}{u,w},   
\end{equation}
with
\begin{equation}
   \mu = p_0\col{\gamma+(1-\gamma)(1-2s) w}+p_1\col{\gamma u w+(1-\gamma)(1-2m) w} +p_2\col{\gamma (1-2m) u+(1-\gamma)}.
\end{equation}

\end{document}